\documentclass[sigconf]{acmart}
\AtBeginDocument{%
  }

\setcopyright{none}
\copyrightyear{2026}
\acmYear{2026}
\acmConference[C\&C '26]{Creativity and Cognition}{July 13--16, 2026}{London, United Kingdom}
\acmBooktitle{Creativity and Cognition (C\&C '26), July 13--16, 2026, London, United Kingdom}
\acmDOI{10.1145/3803784.3809266}
\acmISBN{979-8-4007-2583-8/2026/07}

\begin{document}

\title{World Simulator: Queer Erotica and the Absurdity of AI Video Models That Promise the World}

\author{Adam Cole}
\email{a.cole@arts.ac.uk}
\orcid{0000-0001-9715-314X}
\affiliation{%
  \institution{University of the Arts London}
  \city{London}
  \country{UK}
}

\author{Mick Grierson}
\email{m.grierson@arts.ac.uk}
\orcid{0000-0002-6981-5414}
\affiliation{%
  \institution{University of the Arts London}
  \city{London}
  \country{UK}
}


\begin{abstract}
Increasingly, AI video models are marketed as "world simulators," suggesting their ability to model infinite realities. Despite such claims, these models systematically exclude significant aspects of embodied human experience, particularly sexuality. \textit{World Simulator} is a video installation exploring the poetic friction between these universal claims and the models’ inherent blindness. To do so, the work feeds explicit gay erotica into an AI video-to-video pipeline. Lacking the training data to recognize these images, the system hallucinates surreal alternatives, transforming intimate acts into banal scenes of kitchen appliances, strange architectures, and abstract flesh. By visualizing the limits of synthetic knowledge, the work challenges the hubris of the "world simulator" label, asking how a system, trained primarily on large filtered video datasets, can claim to simulate the world while remaining structurally blind to the body. Beyond this critique, we question the value of simulation itself, asking what forms of sensual representation might offer more expansive, life-affirming possibilities.
\end{abstract}


\begin{teaserfigure}
  \includegraphics[width=\textwidth]{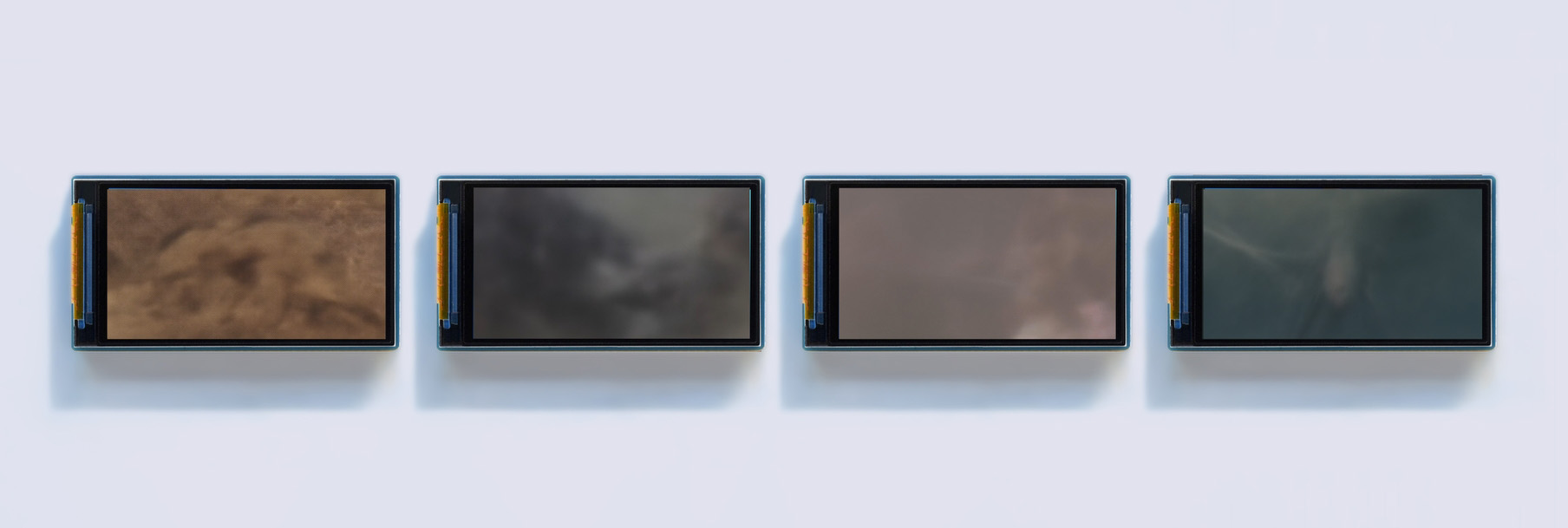}
  \caption{\textbf{World Simulator} install rendering: four miniature screens show an advanced AI video system attempting and failing to reproduce gay erotica, highlighting the limits of knowledge within AI systems sold to us as \textit{world simulators}.}
  \Description{\textbf{World Simulator} install rendering: four screens display an advanced AI video system attempting and failing to reproduce vintage gay erotica, highlighting the limits of knowledge within AI systems sold to us as \textit{world simulators}.}
  \label{fig:teaser}
  \Description{Four miniature, horizontally oriented rectangular screens mounted in a row against a light background. The screens display a sequence of hazy, abstract, earth-toned visuals that include ambiguous, fleshy shapes and darker, undefined textures.}
\end{teaserfigure}


\maketitle

\section{Introduction}
Since 2024, advanced generative video systems have been framed as more than media tools. They are pitched as \textit{world simulators} \cite{bruceGenieGenerativeInteractive2024, openaiVideoGenerationModels2024}. Companies like Google and OpenAI make an ambitious promise: that as video models scale, improving in pixel resolution and physical consistency, they cease to be mere creative engines and become "general purpose simulators of the physical world" \cite{openaiVideoGenerationModels2024}. Moving beyond physics, Google promises that models like Project Genie enable the creation and exploration of "any world imaginable" \cite{bruceGenieGenerativeInteractive2024}. Implicit in this pitch is a claim to completeness: that these systems are building a coherent, totalizing replica of the known world, worthy of our attention and subscription fees.

This project, \textit{World Simulator}, tackles that claim head-on with a simple, provocative question: \textit{what kind of world simulation can exist without sex?}

Currently, commercial large video and world simulator models enforce strict limits on NSFW content \cite{gaoCannotWriteThis2025a, openaiSoraSystemCard2026, runwaymlRunwaysUsagePolicy2026, googleGenerativeAIProhibited2024}. In these digital realities, explicit sexuality (heterosexual and queer alike) is systematically prohibited  \cite{openaiSoraSystemCard2026, runwaymlRunwaysUsagePolicy2026, googleGenerativeAIProhibited2024}. However, passing queer erotica through these sterilized pipelines carries a specific poetic friction, given the historical policing and erasure of queer bodies in traditional media archives \cite{benshoffQueerImagesHistory2005, watsonWhatMattersQueer2024a}. 

This project does not argue that such safety features are inherently bad. Rather, we argue that the total absence of this content highlights the absurdity of the "world simulator" label. Specifically, if these models are simulating "the world," they are simulating a version of it that is structurally blind to one of the most foundational experiences of biological life. The project questions the hubris of presenting these systems as all-seeing and all-knowing, in spite of the fact that they are trained primarily on internet-scale video data \cite{openaiVideoGenerationModels2024, zhuSoraWorldSimulator2024, xingSurveyVideoDiffusion2023}; rather, we should consider whether these models are inherently limited in simulating the complex, visceral, and embodied experience of sexuality and of being human more generally.

Our critical creative AI project provokes this question through a series of tongue-in-cheek video interventions. We take scenes of gay erotica from the classic queer underground film \textit{Boys in the Sand} (1971) \cite{pooleBoysSand1971} by Wakefield Poole, and pass it through a leading video-to-video model. Lacking the "knowledge" to recognize this human behavior, the system fails to reproduce the scene. Instead of a faithful simulation, the erotic choreography is transformed into banal, mundane, or totally abstract representations.

The result is an aesthetic confrontation that visualizes the representational limits of these so-called proto-world-simulators. Furthermore, it challenges the subtext of "simulation" more widely, asking how sensuality could (or should) be represented in synthetic media. In this regard, the work fits squarely within the Creativity \& Cognition theme “Creativity for Change.” It offers an aesthetic experience as a critical intervention, encouraging a skeptical orientation towards AI marketing promises and a deeper reflection on the limits of machine representation.

\begin{figure}[h]
  \centering
  \includegraphics[width=\linewidth]{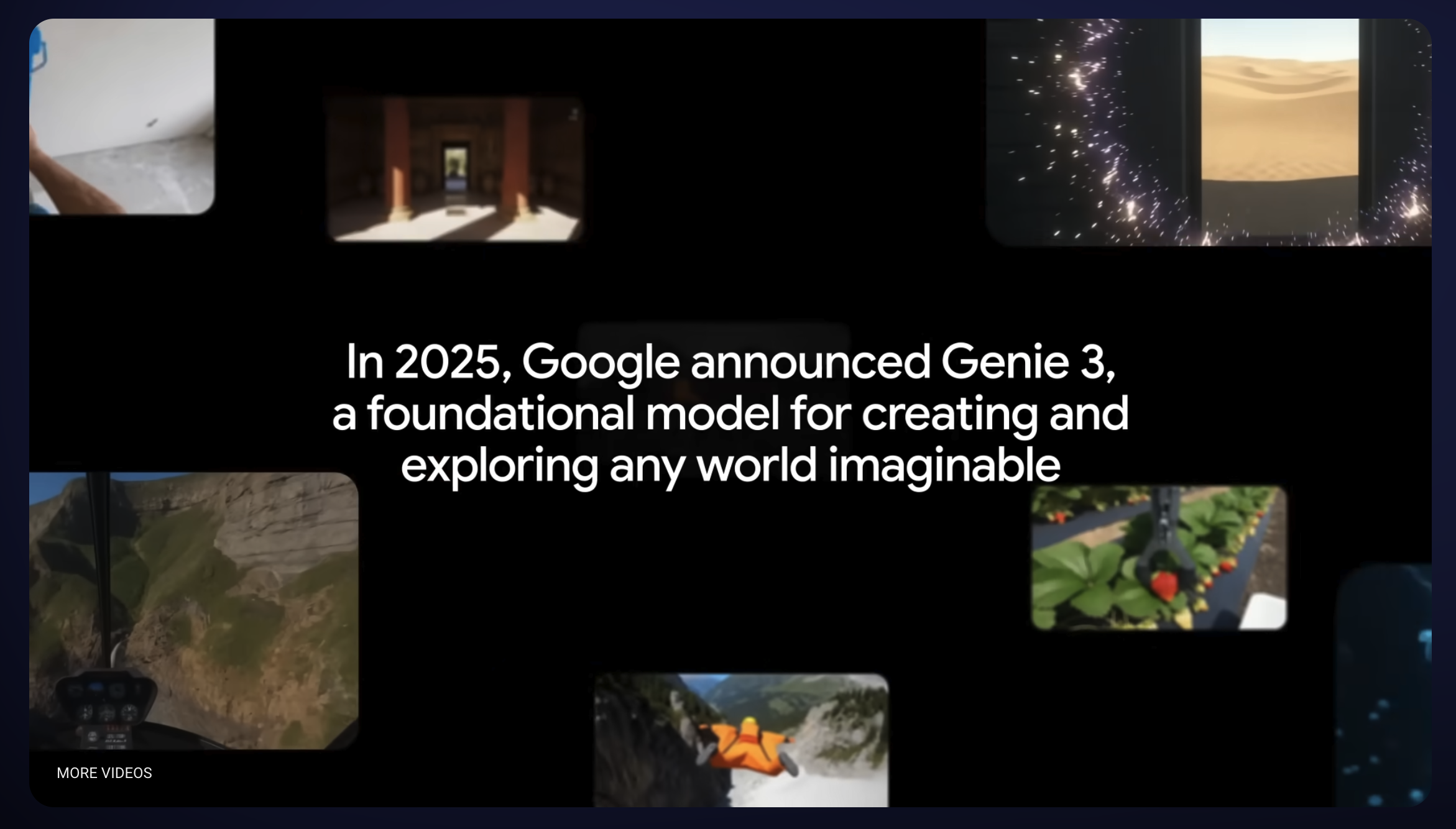}
  \caption{\textbf{Google Genie 3 Promo Video}. Screenshot from promotional video demonstrates the inflated claims of their marketing material. \cite{bruceGenieGenerativeInteractive2024}}
  \label{images/google-promo}
  \Description{A screenshot from a promotional video for Google Genie 3 against a black background. In the center, white text reads: "In 2025, Google announced Genie 3, a foundational model for creating and exploring any world imaginable". Surrounding the text are various thumbnail frames of generated video content, including a person holding a camera, an interior hallway, a desert landscape, an extreme sports athlete skiing, and a close-up of strawberries.}
\end{figure}

\section{Context}

\subsection{Simulation and the Limits of Representation}

There is a rich and long discourse addressing the limits of representation through mediated artifacts, ranging from the shadows of Plato’s cave \cite{platoRepublic1974} to the hyperreality of the postmodernists \cite{baudrillardSimulacraSimulation1994}. This fixation on total simulation is particularly relevant to the tech world and its science fiction fantasies \cite{appioScienceFictionQuest2025}. Inspired by the fabricated worlds of The Matrix and Blade Runner \cite{grbaDeepElseCritical2022}, Silicon Valley has long chased a specific aesthetic dream: a digital reality that is indistinguishable from the physical one \cite{bassettBetterMadeMutual2013, appioScienceFictionQuest2025}.

This ambition has now been accelerated by the discourse on "world simulators". In academic contexts, this term often has a specific, limited meaning: virtual environments that present a coherent physics and causal structure, frequently used for training agents through reinforcement learning \cite{haWorldModels2018, destradeValueguidedActionPlanning2025}. However, in current AI marketing material, the term takes on a more specious turn. OpenAI explicitly frames its video models as precursors to "general purpose simulators of the physical world" \cite{openaiVideoGenerationModels2024}. Meanwhile, Google DeepMind positions tools like Genie not just as creative engines, but as stepping stones toward Artificial General Intelligence (AGI), stating: "While Google DeepMind has a history of agents for specific environments like Chess or Go, building AGI requires systems that navigate the diversity of the real world" \cite{googledeepmindProjectGenieExperimenting2026}.

This project addresses these claims head-on. By highlighting the inability of these models to represent some of the most primal and embodied of human experiences, \textit{World Simulator} exposes the gap between the science-fiction-tinged fantasies of tech marketing and the inherent limits of the technology: limits that may remain technically and ethically insurmountable.


\section{AI Video and Explicit Content}
The discourse around AI and explicit content is vast, spanning ethics \cite{citronSexualPrivacy2019}, law \cite{harrisDeepfakesFalsePornography2019}, and digital safety \cite{gillespieCustodiansInternetPlatforms2018}. Within the realm of closed-source commercial tools (such as Sora \cite{openaiSoraSystemCard2026}, RunwayML \cite{runwaymlRunwaysUsagePolicy2026}, or Veo \cite{googleGenerativeAIProhibited2024}), multiple layers of censorship and safety alignment prevent the generation of nudity or sexual acts. Even in open-source contexts, we can surmise from available research on dataset filtering (such as the massive "cleaning" of the Common Crawl corpus) that explicit content is systematically purged from the training data of foundational models to align with corporate safety standards \cite{dodgeDocumentingLargeWebtext2021}.

\textit{World Simulator} makes no claim about supporting or opposing the inclusion of explicit content in AI models. Rather, it highlights the irony of glorifying a world simulator that possesses no sensual knowledge. The project raises a question about the ontological status of these simulations: what does it mean to simulate a human world while being structurally blind to sex itself?

\subsection{Critical Creative AI}
The field of Critical Creative AI uses machine learning tools not to demonstrate their artistic capabilities, but to critically reflect on their ideological structures and wider social implications \cite{bunzCreativeAIFutures2022, hemmentExperientialAIArts2024}. This practice is exemplified by works such as Parag Mital et al.’s \textit{ Corpus-Based Visual Synthesis} \cite{mitalCorpusbasedVisualSynthesis2013} and Memo Akten et al.’s \textit{Learning to See} \cite{aktenLearningSeeYou2019}, two projects that visualize the structure of neural network comprehension based on their limited training data, exposing the model’s learned priors. In the realm of generative video, Jake Elwes’ \textit{The Zizi Project} disrupts the normative boundaries of deepfake technology by injecting drag performance into the dataset, queering the typically rigid output of GANs \cite{elwesZiziQueeringDataset2019}. Building on this in our previous project, \textit{Me vs. You}, we demonstrate the limits of computer vision systems in interpreting the non-representational qualities of queer intimacy \cite{coleMeVsYou2024}.

\textit{World Simulator} situates itself within this lineage and adopts its interventionist methodology. It operates by confronting the viewer with an aesthetic provocation that renders the abstract limitations of the technology visceral. In doing so, the work transforms a theoretical critique into an active experience.

\section{Video-to-Video Translation as Intervention}
To enact this critical intervention, we utilize the Wan 2.1 VACE (Video All-in-One Creation and Editing) pipeline \cite{wanteamWanOpenAdvanced2025, jiangVACEAllinOneVideo2025}. While not technically a world simulator, it serves as an accessible, state-of-the-art Large Video Model (LVM) \cite{liSurveyLongVideo2024} with a similar architecture to diffusion transformer (DiT) \cite{peeblesScalableDiffusionModels2023b, wanteamWanOpenAdvanced2025} based world simulators \cite{wangMechanisticViewVideo2026, huangVid2WorldCraftingVideo2025}. Crucially, as an open-weight tool, Wan offers the flexibility to bypass the rigid API censorship of commercial platforms, allowing us to process sensitive inputs that would otherwise trigger an immediate refusal. Furthermore, the model demonstrates a "blindness" to sexual content (as seen in our intervention), making it a useful proxy for our wider critique.

\begin{figure}[h]
  \centering
  \includegraphics[width=\linewidth]{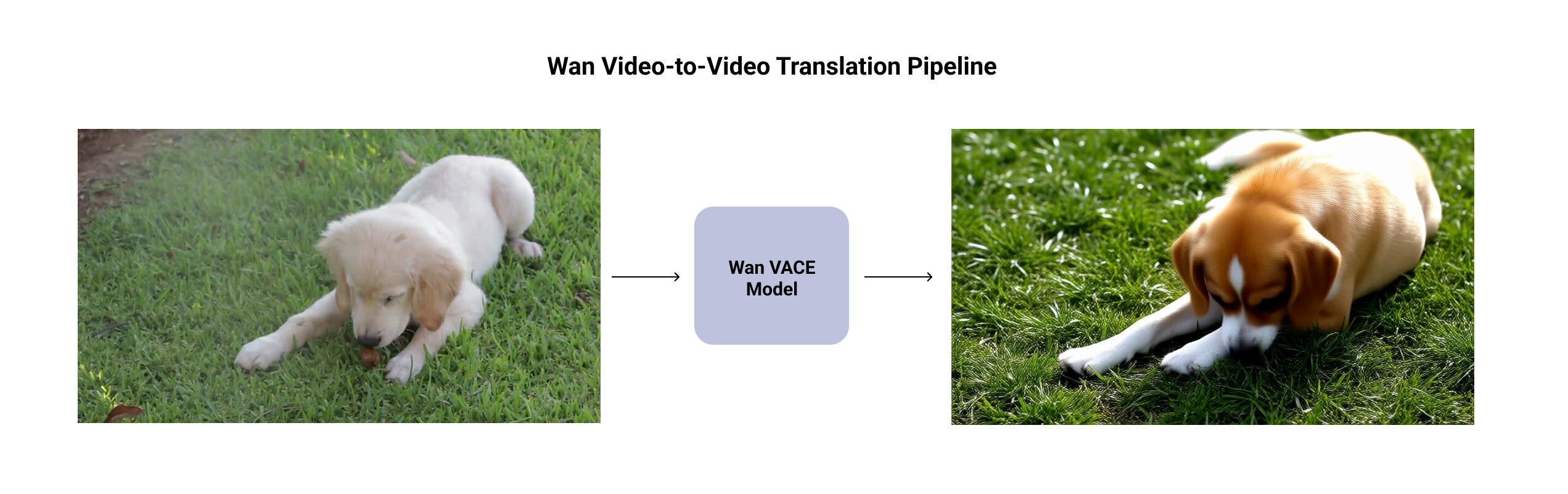}
  \caption{\textbf{Wan VACE Video Translation Pipeline}. The model is able to faithfully reproduce the image of a dog given the source as reference, suggesting a strong learned prior for dog representations. (Dog source image: @Coverr / Pexels, CC0).}
  \label{images/dog-comparison}
  \Description{A schematic diagram illustrating the Wan VACE Video Translation Pipeline. On the left is a source image of a golden retriever puppy lying on green grass. An arrow points to a purple central box labeled "Wan VACE Model". A second arrow points to the output image on the right, which shows a faithful, nearly identical generated image of a dog lying on grass, demonstrating the model's successful semantic translation of a known concept.}
\end{figure}

Our methodology is one of \textit{adversarial translation}. The VACE pipeline is engineered to maintain the temporal consistency of an input video while transforming its style or semantic content (ex Fig.~\ref{images/dog-comparison}). Drawing on the logic of Akten's \textit{Learning to See} \cite{aktenLearningSeeYou2019}, we treat this translation process as a probe into the model's latent space: representations that are easily translated indicate concepts that are well-learned (like the dog in Fig.~\ref{images/dog-comparison}), while inputs that fail to be reproduced (resulting in abstraction or hallucination) reveal the concepts that are effectively "unknown" to the model. By feeding the system content it cannot recognize, we find the boundaries of its so-called "simulation".

\section{Methodology}

To operationalize the adversarial translation methodology outlined above, we curated a dataset of queer video from the erotic film \textit{Boys in the Sand}, selecting clips specifically for their intricate motion and focus on interlocking bodies (features we anticipated the model would struggle to reconstruct).

\begin{figure*}[t]
  \centering
  \includegraphics[width=\textwidth]{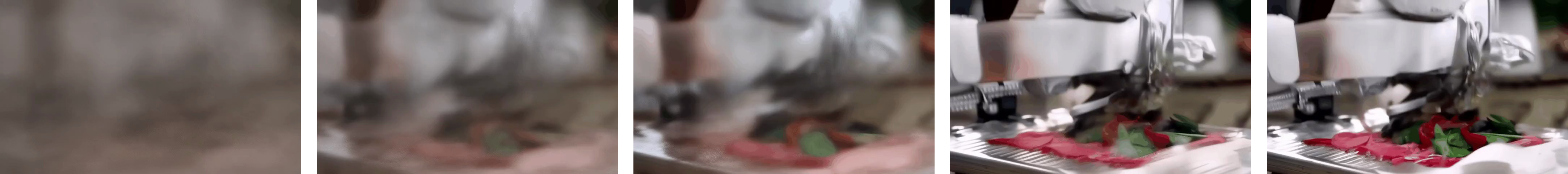}
  \caption{A composite visualizing the generation progression across diffusion steps 1 through 5, tracing the model's rapid resolution from initial "fleshy ambiguity" into a distinct domestic object.}
  \label{images/step-development}
  \Description{A sequential composite of five video frames side-by-side, visualizing the generation progression of an AI model across diffusion steps. From left to right, the screens display images that transition from hazy, unresolved brown and grey smudges into a kitchen appliance, demonstrating the model's collapse from abstract flesh into mundane objects.}
\end{figure*}

Using a custom workflow in ComfyUI \cite{comfy-orgComfyUIMostPowerful2023}, we processed these clips through the Wan 2.1 VACE pipeline. Crucially, rather than using a standard denoising schedule, we executed the generation process with restricted step counts, running distinct iterations at 1, 2, 3, 4, and 5 steps. This manipulation establishes a creative axis from abstraction to representation, allowing us to visualize the model's process across increasing levels of clarity (ex. Fig.~\ref{images/step-development}):

\begin{itemize} 
    \item \textbf{Low Steps (1-2): Ambiguity.} The output is a noisy, hazy blur, but undoubtedly suggestive. The kinetic energy and structural composition of the erotic act are preserved, but the semantic content remains unresolved. 
    \item \textbf{Higher Steps (3-5): Collapse.} As the step count increases, the model forces these ambiguous shapes into known concepts. We observe the "collapse" of the sexual into the mundane, as fleshy shapes harden into machinery or domestic objects. 
\end{itemize}

The resulting outputs were edited into four continuous reels, looping the transition between the suggestive motion of the low-step diffusion and the sterile "correction" of the higher steps. Full generation parameters, such as resolution settings and model checkpoints, are provided in Appendix~\ref{ref:gen-details}.

\begin{figure}[h]
  \centering
  \includegraphics[width=\linewidth]{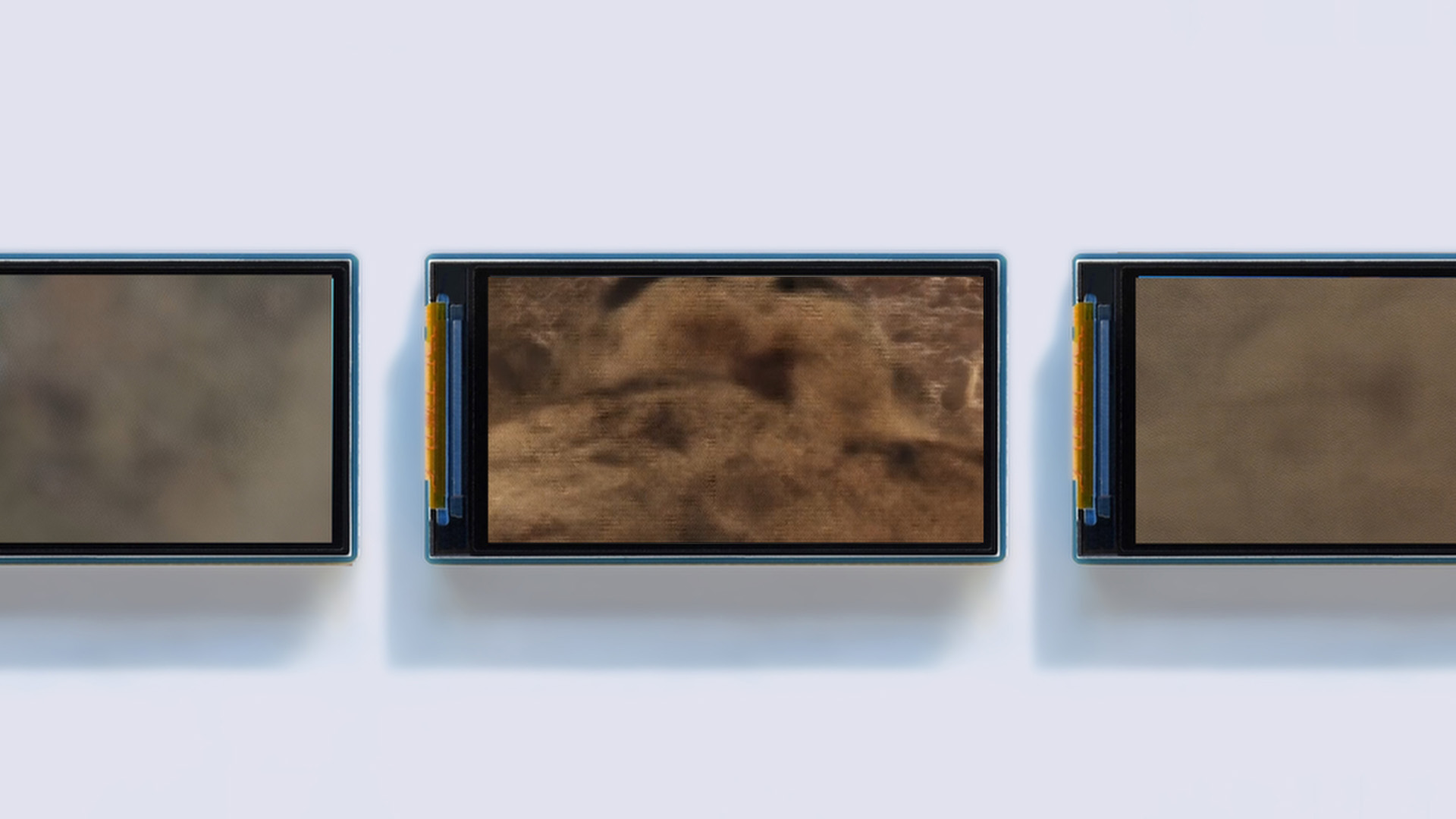}
  \caption{Close-up of a frame representing the mid-diffusion process output, where the visual content remains semantically abstract and ambiguous despite suggestive motion cues.}
  \label{images/single-vid}
  \Description{A close-up view of a single digital screen enclosed in a dark bezel. The screen displays an abstract, blurry image composed of varying shades of brown, tan, and dark shadows. The visual lacks any clearly identifiable objects, representing a state of fleshy ambiguity and unresolved semantic content during the mid-diffusion process.}
\end{figure}

\section{Results}

The final outputs of \textit{World Simulator} are presented in Appendix~\ref{ref:links}. From dozens of generations, we selected sequences that effectively balanced ambiguity with distinct, mundane resolutions. Fig.~\ref{images/single-vid} highlights the output in the middle of the diffusion process, where the erotic motion is visible but the semantic content remains abstract (notably, it is hard to decipher the subtext from the still images alone; the erotic connotation comes primarily from the suggestive motion of the forms). Fig.~\ref{images/step-development} traces the progression across diffusion steps 1 through 5, visualizing the model's transition from fleshy ambiguity to a recognizable domestic object.

For the installation, these clips are edited into four reels displayed on 1.9-inch screens embedded on ESP32 boards.  At this scale, the suggestive composition and motion is legible from cursory viewing without ever becoming outright explicit. The loops emphasize the gap between the input and the output: dwelling on the suggestive 1-step diffusion before revealing the model's banal interpretation. The multi-channel format suggests a system endlessly generating variations, constantly failing to make sense of the erotic world.


\section{Reflection}

\subsection{The False Promise of the World Simulation}

This project demonstrates that the claim of any model being a world simulator is fundamentally flawed, using the extreme example of sex as a provocation. Any model will inherently have blind spots. The training data will always contain biases (whether from curation, censorship, or availability) which propagate through the model and into the outputs of the simulation \cite{benderDangersStochasticParrots2021, nadeemGenderBiasTexttoVideo2025a}. Furthermore, the architecture will always, by design, compress complexity into more predictable representations \cite{crawfordAtlasAIPower2021}.

In this light, \textit{World Simulator} highlights just how shallow claims from Silicon Valley are concerning world simulations. To effectively critique this absurdity, we turn to critical creative practice to more viscerally present our perspective. By visualizing the system's inability to process the erotic, we return to the project's central provocation: can a world simulation really exist without sex?

\subsection{Beyond Simulation}

Crucially, this critique is not a call for foundational models to be trained on pornography for the sake of "completeness." Pornography, too, is a false simulation: a rigid reproduction of tropes and performances that can limit, rather than expand, our understanding of desire.

Instead, we question the utility of "simulation" as an artistic or spiritual goal. We look to the history of queer and experimental cinema, where the objective was rarely the perfect replication of reality, but rather its transcendence. Relevant here is our choice of source material, Poole's \textit{Boys in the Sand}, an underground landmark that elevated queer self-representation by moving beyond the purely pornographic genre \cite{capinoSeminalFantasiesWakefield2004}. It integrates gay erotics with the sublime of the natural environment, prioritizing a dreamlike, sensory expansion over a rigid 'simulation' of sex.

Ironically, in the failure of the machine to parse the body, we find a poetic friction that feels more honest than a perfect rendering. Ultimately, we posit that the pursuit of a total world simulator is a fool’s errand, motivated by spectacle and profit. The role of the artist may not be to help the machine simulate the world more accurately, but to use these tools to create poetic, sensual expansions of our material experience: worlds that do not mimic this one, but offer new ways to feel within it.

\begin{acks}
Adam Cole’s research is supported by the UKRI Techné Studentship, AHRC Grant reference number AH/R01275X/1.
\end{acks}

\bibliographystyle{ACM-Reference-Format}
\bibliography{PhD-bibtex}


\appendix

\section{Video Documentation}
\label{ref:links}
\begin{enumerate}
    \item \textbf{World Simulator Demo}: \url{https://youtu.be/aZPd9cmAmZE}
\end{enumerate}

\section{Video Generation Details}
\label{ref:gen-details}
Below is a sample of the video generation settings used to generate the clips which make up the final work. The generation was run in ComfyUI on an NVIDIA RTX A6000 GPU. 

\begin{verbatim}
generative_pipeline:
  model:
    base: "wan2.1_vace_14B_fp16.safetensors"
    vae: "wan_2.1_vae.safetensors"
    text_encoder: "umt5_xxl_fp16.safetensors"
    
  generation_settings:
    resolution: "832x480"
    frame_count: 49
    sampler: "uni_pc"
    scheduler: "simple"
    steps: 2  # variable between range [1,5]
    cfg_scale: 1.0
    denoise: 1.0
    seed: 845035834986862 # variable

  input_processing:
    source_file: "sample.mp4"  # variable
    start_frame_index: 0 # variable
    control_method: "Canny Edge Detection"
    canny_thresholds:
      low: 0.08
      high: 0.2
\end{verbatim}

\end{document}